\documentclass[prd,twocolumn,superscriptaddress,altaffilletter,showpacs,nofootinbib]{revtex4}
\usepackage[dvips]{graphicx}
\usepackage{amsmath}

\newcommand{\be}{\begin{equation}}
\newcommand{\ee}{\end{equation}}
\newcommand{\bea}{\begin{eqnarray}}
\newcommand{\eea}{\end{eqnarray}}

\begin{document}

\title{Dirac-Born-Infeld Field Trapped in the Braneworld}

\author{Ricardo Garc\'{\i}a-Salcedo}\email{rigarcias@ipn.mx}\affiliation{Centro de Investigacion en Ciencia Aplicada y Tecnologia Avanzada - Legaria
del IPN, M\'exico D.F., M\'exico.}

\author{Dania Gonzalez}\email{dgm@uclv.edu.cu}\affiliation{Departamento de Matem\'{a}tica, Universidad Central de Las Villas, 54830
Santa Clara, Cuba.}

\author{Tame Gonzalez}\email{tamegc72@gmail.com}\affiliation{Departamento de F{\'\i}sica, Centro de Investigaci\'on y de Estudios
Avanzados del IPN, A.P. 14-740, 07000 M\'exico D.F., M\'exico.}

\author{Claudia Moreno}\email{claudia.moreno@cucei.udg.mx}\affiliation{Departamento de F\'{\i}sica y Matem\'aticas, Centro Universitario de
Ciencias Ex\'actas e Ingenier\'{\i}as, Corregidora 500 S.R., Universidad de Guadalajara, 44420 Guadalajara, Jalisco, M\'exico.}

\author{Israel Quiros}\email{israel@uclv.edu.cu}\affiliation{Divisi\'on de Ciencias e Ingenier\'ia de la Universidad de Guanajuato, A.P.
150, 37150, Le\'on, Guanajuato, M\'exico.}

\date{\today}

\begin{abstract}
We apply the dynamical systems tools to study the (linear) cosmic dynamics of a Dirac-Born-Infeld-type field trapped in the braneworld. We focus,
exclusively, in Randall-Sundrum and in Dvali-Gabadadze-Porrati brane models. We analyze the existence and stability of asymptotic solutions for the AdS throat and the quadratic potential and a particular choice of the warp factor and of the potential for the DBI field ($f(\phi)=1/V(\phi)$). It is demonstrated, in particular, that in the ultra-relativistic approximation matter-scaling and scalar field-dominated solutions always arise. In the first scenario the empty universe is the past attractor, while in the second model the past attractor is the matter-dominated phase.
\end{abstract}

\maketitle

\section{Introduction}

Recent observations from the Wilkinson Microwave Anisotropy Probe (WMAP)
\cite{bennet} offer strong supporting evidence in favor of the inflationary
paradigm. In the most simple models of this kind, the energy density of the
universe is dominated by the potential energy of a single (inflaton) scalar
field that slowly rolls down in its self-interaction potential \cite%
{starobinsky}. Restrictions imposed upon the class of potentials which can
lead to realistic inflationary scenarios, are dictated by the slow-roll
approximation, and hence, the result is that only sufficiently flat
potentials can drive inflation. In order for the potential to be
sufficiently flat, these conventional inflationary models should be
fine-tuned. This simple picture of the early-time cosmic evolution can be
drastically changed if one considers models of inflation inspired in
``Unified Theories'' like the Super String or M-theory. The most appealing
models of this kind are the Randall-Sundrum braneworld model of type 2 (RS2)
\cite{rs} and Dvali-Gabadadze-Porrati (DGP) brane worlds \cite{dgp}.

In the RS2 model a single co-dimension 1 brane with positive tension is
embedded in a five-dimensional anti-de Sitter (AdS) space-time, which is
infinite in the direction perpendicular to the brane. In general, the
standard model particles are confined to the brane, meanwhile gravitation
can propagate in the bulk. In the low-energy limit, due to the curvature of
the bulk, the graviton is confined to the brane, and standard
(four-dimensional) general relativity laws are recovered. RS2 braneworld
models have an appreciable impact on early universe cosmology, in
particular, for the inflationary paradigm. In effect, a distinctive feature
of cosmology with a scalar field confined to a RS2 brane is that the
expansion rate of the universe differs at high energy from that predicted by
standard general relativity. This is due to a term -- quadratic in the
energy density -- that produces enhancing of the friction acting on the
scalar field. This means that, in RS2 braneworld cosmology, inflation is
possible for a wider class of potentials than in standard cosmology \cite%
{hawkins}. Even potentials that are not sufficiently flat from the point of
view of the conventional inflationary paradigm can produce successful
inflation. At sufficiently low energies (much less than the brane tension),
the standard cosmic behavior is recovered prior to primordial
nucleosynthesis scale ($T\sim 1\; MeV$) and a natural exit from inflation
ensues as the field accelerates down its potential \cite{5}.\footnote{%
In this scenario, reheating arises naturally even for potentials without a
global minimum and radiation is created through gravitational particle
production \cite{6} and/or through curvaton reheating \cite{7}. This last
ingredient improves the brane ``steep'' inflationary picture \cite{8}. Other
mechanisms such as preheating, for instance, have also been explored \cite{9}%
.} The DGP model describes a brane with 4D world-volume, that is embedded
into a flat 5D bulk, and allows for infrared (IR)/large scale modifications
of gravitational laws. A distinctive ingredient of the model is the induced
Einstein-Hilbert action on the brane, that is responsible for the recovery
of 4D Einstein gravity at moderate scales, even if the mechanism of this
recovery is rather non-trivial \cite{deffayet}. Nevertheless, studying the
dynamics of DGP models continues being a very attractive subject of research
\cite{quiros}. It is due, in part, to the very simple geometrical
explanation to the \textquotedblleft dark energy" problem and the fact that
it is one of a very few possible consistent IR modifications of gravity that
might be ever found. The acceleration of the expansion at late times is
explained here as a consequence of the leakage of gravity into the bulk at
large (cosmological) scales, which is just a 5D geometrical effect.

Nonlinear scalar field theories of the Dirac-Born-Infeld (DBI) type have
attracted much attention in recent years due to their role in models of
inflation based on string theory. DBI inflation \cite%
{speedlimit,chen,basic,DBInflation,chen1} is motivated by brane inflationary
models \cite{dvali} in warped compactifications \cite{verlinde}. These
scenarios identify the inflaton with the position of a mobile D-brane moving
on a warped (compact) 6-dimensional submanifold of spacetime (for reviews
and references see \cite{string}), which means that the inflaton is
interpreted as an open string mode. In these models, the warped space slows
down the rolling of the inflaton on even a steep potential, making easier
inflation. This slowing down can also be understood as arising due to
interactions between the inflaton and the strongly coupled large-N dual
field theory \cite{chen1}. This scenario can naturally arise in warped
string compactifications \cite{chen}. Usually only effective
four-dimensional DBI cosmological models are studied.

It is our opinion that studying the impact higher-dimensional brane effects
have on the cosmic dynamics of DBI-type models, is a task worthy of
attention. A DBI field trapped in a RS brane could be a nice scenario to
make early-time inflation easier, both, because of the interaction of the
inflaton with the strongly coupled (large-N) dual field theory, and because
of the UV brane effects. Meanwhile a DBI scalar field confined to a
(self-accelerating) DGP braneworld could be a useful arena where to address
unified description of early-time inflation, fueled by the ``slowing down''
effect of the warped space, and late-time speed-up, originated from UV
leakage of gravity into the extra-space.

Aim of this paper is, precisely, to investigate the dynamics of a DBI-type
field trapped in a Randall-Sundrum brane and in a Dvali-Gabadadze-Porrati
braneworld, respectively, by invoking the dynamical systems tools. The study
of the asymptotic properties of these models allows to correlate such
important dynamical systems concepts like past and future attractors -- as
well as saddle equilibrium points -- with generic cosmological evolution. In
a sense the present work might be considered as a natural completion of the
one reported in Ref. \cite{basic}. For this reason, as in the above
reference, here we concentrate in the case of an anti-de Sitter (AdS) throat
and quadratic self-interaction potential for the inflaton and a particular choice of the warp factor and of the potential for the DBI field ($f(\phi)=1/V(\phi)$). In addition to the scalar field we also consider a background fluid trapped on the
braneworld. Through the paper we use natural units ($8\pi
G=8\pi/m_{Pl}^2=\hbar=c=1$).

\section{DBI Action}

In the region where the back-reaction and stringy physics can be ignored,
the effective action for the DBI field has the following form \cite{chen1}:

\begin{eqnarray}
S_{DBI}=-\int d^4x \sqrt{|g|}\{f^{-1}(\phi)\sqrt{1+f(\phi)(\nabla\phi)^2}
\notag \\
-f^{-1}(\phi)+V(\phi)\},  \label{dbi}
\end{eqnarray}
where $\phi$ is the inflaton, $V(\phi)$ - its self-interaction potential,
and $f(\phi)$ is the warping factor. For a spatially flat FRW metric $%
(\nabla\phi)^2=-\dot\phi^2$, where the dot accounts for derivative in
respect to the cosmic time. The equation of motion for the DBI inflaton $\phi
$ can be written in the following way:

\begin{eqnarray}
\ddot\phi+\frac{3\partial_\phi f}{2f}\dot\phi^2-\frac{\partial_\phi f}{f^2}%
+3\gamma_L^{-2}H\dot\phi  \notag \\
+\gamma_L^{-3}\left(\partial_\phi V+\frac{\partial_\phi f}{f^2}\right)=0,
\label{dbikg}
\end{eqnarray}
where the ``Lorentz boost'' $\gamma_L$ is defined as

\begin{equation}
\gamma_L=\frac{1}{\sqrt{1-f(\phi)\dot\phi^2}}.  \label{lfactor}
\end{equation}

Alternatively the equation of motion of the DBI-type field can be written in
the form of a continuity equation:

\begin{equation}
\dot\rho_\phi+3H(\rho_\phi+p_\phi)=0,  \label{dbicontinuity}
\end{equation}
where we have defined the following energy density and pressure of the DBI
scalar field:

\begin{equation}
\rho_\phi=\frac{\gamma_L-1}{f}+V(\phi),\;\;p_\phi=\frac{\gamma_L-1}{\gamma_L
f}-V(\phi),  \label{definitions}
\end{equation}
respectively. In this paper we concentrate just  in two case: i) a AdS throat which
amounts to consider $f(\phi)=\alpha/\phi^4$, where $\alpha$ in specific
string constructions is a parameter which depends on the flux numbers \cite%
{chen1},\footnote{%
In general, inflationary observables may depend on the details of the warp
factor \cite{maiden}, however, if we assume that the last 60 e-foldings of
inflation occur far from the tip of the throat, the above is a good
approximation.} and a quadratic self-interaction potential $%
V(\phi)=m^2\phi^2/2$ and ii) a particular choice of the warp factor and of the potential for the DBI field ($f(\phi)=1/V(\phi)$).

\section{Autonomous Systems}

The dynamical systems tools offer a very useful approach to the study of the
asymptotic properties of the cosmological models \cite{coley}. In order to
be able to apply these tools one has to follow these steps: i) to identify
the phase space variables that allow writing the system of cosmological
equations in the form of an autonomus system of ordinary differential
equations (ODE).\footnote{%
There can be several different possible choices, however, not all of them
allow for the minimum possible dimensionality of the phase space.}, ii) with
the help of the chosen phase space variables, to build an autonomous system
of ODE out of the original system of cosmological equations, and iii) (a
some times forgotten or under-appreciated step) to indentify the phase space
spanned by the chosen variables, that is relevant to the cosmological model
under study. After this one is ready to apply the standard tools of the
(linear) dynamical systems analysis.

The goal of the dynamical systems study is to correlate such important
concepts like past and future attractors (also, saddle critical points) with
asymptotic cosmological solutions. If a given cosmological solution can be
associated with a critical point in the phase space of the model, this means
that, independent of the initial data, the universe's dynamics will evolve
for a sufficiently long time in the neightbourhood of this solution,
otherwise, it represents a quite generic phase of the cosmic dynamics.

In the following subsections we keep the expressions as general as possible,
and then, in section IV we substitute the above mentioned expressions for $%
f(\phi)$ and $V(\phi)$.

\subsection{The DBI-RS Model}

Here we will be concerned with the dynamics of a DBI inflaton that is
trapped in a Randall-Sundrum brane of type 2. The field equations -- in
terms of the Friedmann-Robertson-Walker (FRW) metric -- are the following:

\begin{eqnarray}
&&3H^{2}=\rho _{T}(1+\frac{\rho _{T}}{2\lambda }),  \label{FriedRS} \\
&&2\dot{H}=-(1+\frac{\rho _{T}}{\lambda })\left[ \gamma _{L}\dot{\phi}%
^{2}+(1+\omega _{m})\rho _{m}\right],  \label{rsdbi}
\end{eqnarray}
where $\omega _{m}$ is the equation of state (EOS) parameter of the matter
fluid, while $\rho _{T}=\rho _{\phi }+\rho _{m}$ -- the total energy density
on the brane. Additionally one has to consider the continuity equations for
the DBI-type field (equation (\ref{dbikg})) or, alternatively, (\ref%
{dbicontinuity})) and for the matter fluid:

\begin{equation}
\dot\rho_m+3(1+\omega_m)H\rho_m=0.  \label{mattercontinuity}
\end{equation}
The model described by the above equations will be called as ``DBI-RS
model''. Our aim will be to write the latter system of second-order
(partial) differential equations, as an autonomous system of (first order)
ordinary differential equations. For this purpose we introduce the following
phase variables \cite{basic}:

\begin{eqnarray}
&&x\equiv\frac{1}{H}\sqrt{\frac{\gamma_L}{3f}},\;y\equiv\frac{\dot\phi\sqrt{%
\gamma_L}}{H},\;z\equiv\frac{\sqrt{V}}{\sqrt{3}H},\;r\equiv\frac{\rho_T}{3H^2%
},  \notag \\
&&\mu_1\equiv\frac{\partial_\phi V}{V^{3/2}f^{1/2}},\;\mu_2\equiv\frac{%
\partial_\phi f}{V^{3/2}f^{5/2}}.  \label{phase var}
\end{eqnarray}
It can be realized that, in terms of the variable $r$,

\begin{equation}
\frac{\rho_T}{\lambda}=\frac{2(1-r)}{r},\;\Rightarrow\;0<r\leq 1.
\label{zconstraint}
\end{equation}
This means that the four-dimensional (low-energy) limit of the
Randall-Sundrum cosmological equations -- corresponding to the formal limit $%
\lambda\rightarrow\infty$ -- is associated with the value $r=1$. The
high-energy limit $\lambda\rightarrow 0$, on the contrary, corresponds to $%
r\rightarrow 0$. We write the Lorentz boost in terms of the variables of
phase space as:

\begin{equation}
\gamma\equiv\frac{1}{\gamma_L}=\sqrt{1-\frac{y^2}{3x^2}}.  \label{lboost}
\end{equation}
Standard "non-relativistic" behavior corresponds to $\gamma=1$, while the
"ultra-relativistic" (UR) regime is associated with $\gamma=0$. In terms of
the variables that span the phase space ($x,y,z,r,\mu_1,\mu_2$), the
cosmological equations (\ref{dbicontinuity}), (\ref{FriedRS}), (\ref{rsdbi}%
), and (\ref{mattercontinuity}), can be written as an autonomous system of
ordinary differential equations (ODE):

\begin{eqnarray}
&&x^{\prime }=-\frac{1}{2}(\mu _{1}+\mu _{2})\frac{yz^{3}}{x^{2}}-\frac{y^{2}%
}{2x}-x\frac{H^{\prime }}{H},  \label{eqx} \\
&&y^{\prime }=-\frac{3}{2}[\mu _{1}(\gamma ^{2}+1)+\mu _{2}(\gamma -1)^{2}]%
\frac{z^{3}}{x}  \notag \\
&&\;\;\;\;\;\;\;\;\;\;\;\;\;\;\;\;\;\;\;\;\;\;\;\;\;\;\;-\frac{3}{2}(\gamma
^{2}+1)y-y\frac{H^{\prime }}{H},  \label{eqy} \\
&&z^{\prime}=\frac{1}{2}\mu _{1}\frac{yz^{2}}{x}-z\frac{H^{\prime}}{H}%
,\;\;r^{\prime}=\frac{2r(r-1)}{2-r}\frac{H^{\prime }}{H},  \label{eqzr} \\
&&\mu_{1}^{\prime}=\mu_{1}^{2}\frac{yz}{x}\left[\Gamma _{V}-\frac{3}{2}-%
\frac{\partial _{\phi }\ln f}{\partial _{\phi }\ln V^{2}}\right],
\label{eqm1} \\
&&\mu_{2}^{\prime }=\frac{\mu_{2}^{2}}{\gamma}\frac{yz^{3}}{x^{3}}\left[%
\Gamma _{f}-\frac{5}{2}-\frac{\partial _{\phi }\ln V^{3}}{\partial _{\phi
}\ln f^{2}}\right],  \label{eqm2}
\end{eqnarray}
where the prime denotes derivative with respect to the number of e-foldings $%
\tau\equiv\ln a$, while
\begin{equation*}
\Gamma_{V}\equiv\frac{(V\partial _{\phi}^{2}V)}{(\partial_{\phi}V)^{2}}%
,\;\;\Gamma_{f}\equiv\frac{(f\partial_{\phi}^{2}f)}{(\partial_{\phi}f)^{2}},
\end{equation*}
and:

\begin{equation}
\frac{H^{\prime}}{H}=-\frac{2-r}{2r}\left\{y^2+3(\omega_m+1)\left[%
r-(1-\gamma)x^2-z^2\right]\right\}.  \label{h'/h}
\end{equation}

It will be helpful to have the parameters of observational importance $%
\Omega_\phi=\rho_\phi/3H^2$ -- the scalar field dimensionless energy density
parameter, and the equation of state (EOS) parameter $\omega_\phi=p_\phi/%
\rho_\phi$, written in terms of the variables of phase space:

\begin{equation}
\Omega_\phi=(1-\gamma)x^2+z^2,\;\;\omega_\phi=\frac{(1-\gamma)\gamma x^2-z^2%
}{(1-\gamma)x^2+z^2}.
\end{equation}
Recall, also, that the deceleration parameter $q=-(1+H^{\prime}/H)$:
\begin{equation*}
q=-1+\frac{2-r}{2r}\{y^2+3(\omega_m+1)[r-(1-\gamma)x^2-z^2]\}.
\end{equation*}

\subsection{DBI-DGP model}

In this section we focus our attention in a braneworld model where the DBI
inflaton is confined to a DGP brane. In the (flat) FRW metric, the
cosmological (field) equations are the following:

\begin{eqnarray}
&&Q_{\pm}^{2}=\frac{1}{3}(\rho_{m}+\rho_{\phi}),  \notag \\
&&\dot{\rho}_{m}=-3(1+\omega_{m})H\rho_{m},  \label{fieldequation}
\end{eqnarray}
where, as before $\omega _{m}$ is the EOS parameter of the matter fluid, $%
\rho _{m}$ is the energy density of the background barotropic fluid and $%
\rho _{\phi }$ is the energy density of DBI field. Also one has to consider
the continuity equations for the DBI-type field (equation (\ref{dbikg}) or,
alternatively, equation (\ref{dbicontinuity})). We have used the following
definition:

\begin{equation}
Q_\pm^2\equiv H^2\pm \frac{1}{r_c}H,  \label{Qequation}
\end{equation}
with $r_c$ being the so called crossover scale. In what follows we will
refer to this model as the "DBI-DGP model". There are two possible branches
of the DGP model corresponding to the two possible choices of the signs in (%
\ref{Qequation}): "+" is for the normal DGP models that are free of ghost,
while "-" is for the self-accelerating solution. As before, our goal will be
to write the latter system of second-order (partial) differential equations,
as an autonomous system of (first order) ordinary differential equations.
For this purpose we introduce the following phase variables:

\begin{equation}
x\equiv\frac{1}{Q}\sqrt{\frac{\gamma_{L}}{3f}},\;y\equiv\frac{\dot{\phi}%
\sqrt{\gamma_{L}}}{Q},\;z\equiv\frac{\sqrt{V}}{\sqrt{3}Q},\;r\equiv\frac{Q}{H%
},  \label{var}
\end{equation}
The expression determining the Lorentz boost coincides with (\ref{lboost}):
\begin{equation*}
\gamma\equiv\frac{1}{\gamma_{L}}=\sqrt{1-\frac{y^{2}}{3x^{2}}}.
\end{equation*}
Hence, as before, $\gamma=1$ is for the non-relativistic case, while $%
\gamma=0$ is for the UR regime.

After the above choice of phase space variables the cosmological equations
can be written as an autonomous system of ODE:

\begin{eqnarray}
&&x^{\prime }=-\frac{y}{2x^{2}}\left[ xy+z^{3}r(\mu _{1}+\mu _{2})\right] -x%
\frac{Q^{\prime }}{Q}.  \label{eqx'} \\
&&y^{\prime }=-\frac{3z^{3}r}{2x}\left[ \mu _{1}(\gamma
^{2}+1)+\mu_{2}(\gamma -1)^{2}\right] -  \notag \\
&&\;\;\;\;\;\;\;\;\;\;\;\;\;\;\;\;\;\;\;\;\;\;\;\;\;\;\;\;-\frac{3y}{2}%
(\gamma ^{2}+1)-y\frac{Q^{\prime }}{Q},  \label{eqy'} \\
&&z^{\prime }=\frac{yz^{2}r}{2x}\mu _{1}-z\frac{Q^{\prime }}{Q}%
,\;\;r^{\prime }=r\left( \frac{1-r^{2}}{1+r^{2}}\right)\frac{Q^{\prime}}{Q}.
\label{eqzr'}
\end{eqnarray}
Recall that the prime denotes derivative with respect to the number of
e-foldings $\tau\equiv\ln a$, while
\begin{equation*}
\mu _{1}\equiv \frac{\partial_{\phi }V}{V^{3/2}f^{1/2}},\;\mu_{2}\equiv\frac{%
\partial_{\phi}f}{V^{3/2}f^{5/2}}.
\end{equation*}
We have considered also the following relationship:
\begin{equation*}
\frac{H^{\prime}}{H}=\frac{2r^{2}}{1+r^{2}}\frac{Q^{\prime }}{Q},
\end{equation*}
where

\begin{equation}
\frac{Q^{\prime }}{Q}\equiv -\frac{1}{2}\{3(1+\omega
_{m})[1-x^{2}(1-\gamma)-z^{2}]+y^{2}\}.  \label{q'/q}
\end{equation}
Equations (\ref{eqx'})-(\ref{eqzr'}) have to be complemented with the
addition of equations (\ref{eqm1}) and (\ref{eqm2}) above, which are the
autonomous ordinary differential equations for $\mu_{1}$ and $\mu_{2}$,
respectively. The equation (\ref{Qequation}) can be rewritten as:

\begin{equation}
r^2=1\pm\frac{1}{r_c H}.  \label{Qequationr}
\end{equation}

For the Minkowski phase, since $0\leq H\leq \infty $ (we consider just
non-contracting universes), then $1\leq r\leq \infty $. The case $-\infty
\leq r\leq -1$ corresponds to the time reversal of the latter situation. For
the self-accelerating phase, $-\infty \leq r^{2}\leq 1$, but since we want
real valued $r$ only, then $0\leq r^{2}\leq 1$.\footnote{%
In fact, fitting SN observations requires $H\geq r_{c}^{-1}$ in order to
achieve late-time acceleration (see, for instance, reference \cite{koyama}
and references therein). This means that $r$ has to be real-valued.} As
before, the case $-1\leq r\leq 0$ represents time reversal of the case $%
0\leq r\leq 1$ that will be investigated here.\footnote{%
Points with $r=0$ and their neighborhood have to be carefully analyzed due
to the fact that at $r=0$, other phase space variables (see Eq.(\ref{var}))
and equations in (\ref{eqx'}-\ref{eqzr'}) are undefined in general.} Both
branches share the common subset $(x,y,z,r=1)$, which corresponds to the
formal limit $r_{c}\rightarrow \infty $ (see equation (\ref{Qequation})), i.
e., this represents just the standard behavior typical of Einstein-Hilbert
theory coupled to a self-interacting scalar field.

To finalize this section we write useful magnitudes of observational
interest in terms of the dynamical variables (\ref{var}). The ``effective''
dimensionless density parameters $\bar{\Omega}_m=\rho_m/3Q^2$ and $\bar{%
\Omega}_\phi=\rho_\phi/3Q^2$:

\begin{eqnarray}
&&\bar{\Omega}_{m}=\frac{\Omega_{m}}{r^{2}}=1-x^{2}(1-\gamma)-z^{2},  \notag
\\
&&\bar{\Omega}_{\phi}=\frac{\Omega_{\phi}}{r^{2}}=x^{2}(1-\gamma)+z^{2}.
\label{omega}
\end{eqnarray}
For the equation of state (EOS) $\omega_{\phi}=p_{\phi}/\rho_{\phi}$:

\begin{equation}
\omega_\phi=\frac{x^2(1-\gamma)\gamma-z^2}{x^2(1-\gamma)+z^2}.  \label{EOS}
\end{equation}

\subsection{Brane Effects in the Phase Space}

Higher-dimensional (brane) effects are encrypted in the variable $r$ for the
DBI-RS model as well as for the DBI-DGP model. In both cases the
hyper-surfaces $\sum$ in the phase space, that are foliated by the value $r=1
$: $\sum=\sum(x,y,z,r=1,\mu_{1},\mu_{2})$, represent the loci of the
equilibrium configurations that can be associated with standard general
relativity behavior. For the DBI-DGP model the phase space hyper-plane $\sum$
represents, additionally, the boundary in the phase space separating the
Minkowski cosmological phase ($r\geq 1$) from the the self-accelerating one (%
$0\leq r\leq 1$). Trajectories in the phase space that scape from $\sum$ are
associated with higher-dimensional modifications of general relativity
produced by the brane effects.

\begin{table*}[tbp]
\caption[crit]{Properties of the critical points for the autonomous system (%
\protect\ref{asode1}).}
\label{tab1}%
\begin{tabular}{@{\hspace{4pt}}c@{\hspace{14pt}}c@{\hspace{14pt}}c@{\hspace{14pt}}c@{\hspace{14pt}}c@{\hspace{14pt}}c@{\hspace{14pt}}c@{\hspace{14pt}}c@{\hspace{14pt}}c@{\hspace{14pt}}c@{\hspace{14pt}}c}
\hline\hline
&  &  &  &  &  &  &  &  &  &  \\[-0.3cm]
Equilirium point & $x$ & $y$ & $z$ & $r$ & Existence & $\gamma$ & $\Omega_m$
& $\Omega_\phi$ & $\omega_\phi$ & $q$ \\[0.1cm] \hline
&  &  &  &  &  &  &  &  &  &  \\[-0.2cm]
$E$ & $0$ & $0$ & $0$ & $0$ & Always & Undefined & $0$ & $0$ & Undefined & $%
3\omega_m+2$ \\[0.2cm]
$M$ & $0$ & $0$ & $0$ & $1$ & " & Undefined & $1$ & $0$ & Undefined & $%
(3\omega_m+1)/2$ \\[0.2cm]
$U^\pm$ & $1$ & $\pm\sqrt{3}$ & $0$ & $1$ & " & $0$ & $0$ & $1$ & $0$ & $1/2$
\\[0.2cm] \hline\hline
\end{tabular}%
\end{table*}

\begin{table*}[tbp]
\caption[eigenv]{Eigenvalues of the linearization matrices corresponding to
the critical points in table \protect\ref{tab1}.}
\label{tab2}%
\begin{tabular}{@{\hspace{4pt}}c@{\hspace{14pt}}c@{\hspace{14pt}}c@{\hspace{14pt}}c@{\hspace{14pt}}c}
\hline\hline
&  &  &  &  \\[-0.3cm]
Equilibrium point & $\lambda _{1}$ & $\lambda _{2}$ & $\lambda _{3}$ & $%
\lambda _{4}$ \\[0.1cm] \hline
&  &  &  &  \\[-0.2cm]
$E$ & $3\omega _{m}$ & $3(\omega _{m}+1)$ & $3(\omega _{m}+1)$ & $%
3(\omega_{m}+1)$ \\[0.2cm]
$M$ & $-3(1+\omega_{m})$ & $3(\omega_{m}-1)/2$ & $3(\omega_{m}+1)/2$ & $%
(3\omega_{m}+1)/2$ \\[0.2cm]
$U^{\pm}$ & $-3\omega_{m}$ & $-3$ & $3/2$ & $3$ \\[0.2cm] \hline\hline
\end{tabular}%
\end{table*}

\begin{table*}[tbp]
\caption[crit]{Properties of the equilibrium points of the autonomous system
(\protect\ref{ultra1}). Here $\protect\gamma_{m}=\protect\omega_{m}+1$,
while $\protect\eta\equiv\protect\mu(\protect\sqrt{\protect\mu^{2}+12}-%
\protect\mu)$.}
\label{tab1'}%
\begin{tabular}{@{\hspace{4pt}}c@{\hspace{14pt}}c@{\hspace{14pt}}c@{\hspace{14pt}}c@{\hspace{14pt}}c@{\hspace{14pt}}c@{\hspace{14pt}}c@{\hspace{14pt}}c@{\hspace{14pt}}c}
\hline\hline
&  &  &  &  &  &  &  &  \\[-0.3cm]
Equilibrium point & $y$ & $z$ & $r$ & Existence & $\Omega _{m}$ & $%
\Omega_{\phi }$ & $\omega _{\phi }$ & $q$ \\ \hline
&  &  &  &  &  &  &  &  \\
$E$ & $0$ & $0$ & $0$ & Always & $0$ & $0$ & Undefined & $3\omega _{m}+2$ \\%
[0.2cm]
$M$ & $0$ & $0$ & $1$ & " & $1$ & $0$ & Undefined & $\frac{3\omega _{m}+1}{2}
$ \\[0.2cm]
$K$ & $\pm\sqrt{3}$ & $0$ & $1$ & " & $0$ & $1$ & $0$ & $\frac{1}{2}$ \\%
[0.2cm]
$SF$ & $\pm\sqrt{\frac{\eta}{2}}$ & $\frac{\eta}{(2\sqrt{3}\mu)}$ & $1$ & $%
\mu\geq 0$ & $0$ & $1$ & $-1+\frac{\eta}{6}$ & $-1-\frac{\eta}{4}$ \\[0.2cm]
$MS$ & $\pm\frac{3\gamma_{m}}{\mu}\sqrt{-\frac{\gamma_{m}}{\omega_{m}}}$ & $%
\frac{\sqrt{3}\gamma_{m}}{\mu}$ & $1$ & $-1<\omega_{m}<0$ & $1+\frac{%
3\gamma_{m}^{2}}{\mu^{2}\omega_{m}}$ & $-\frac{3\gamma_{m}^{2}}{%
\mu^{2}\omega_{m}}$ & $\frac{\omega_{m}}{2\omega_{m}+1}$ & $\frac{%
3\omega_{m}+1}{2}$ \\[0.2cm] \hline\hline
\end{tabular}
\newline
\end{table*}

\begin{table*}[tbp]
\caption[eigenv]{Eigenvalues of the linearization matrices corresponding to
the equilibrium points in table \protect\ref{tab1'}. The eigenvalues
corresponding to the fourth point ($SF$) in Tab.\protect\ref{tab1'}, have
not been included due to their overwhelming complexity. Here $\protect\xi%
\equiv\protect\sqrt{\protect\mu^2(9\protect\omega_m^2+6\protect\omega%
_m+1)+24(\protect\omega_m^3+3\protect\omega_m^2+3\protect\omega_m+1)}$.}
\label{tab2'}%
\begin{tabular}{@{\hspace{4pt}}c@{\hspace{14pt}}c@{\hspace{14pt}}c@{\hspace{14pt}}c@{\hspace{14pt}}c}
\hline\hline
&  &  &  &  \\[-0.3cm]
Equilibrium point & $\lambda_1$ & $\lambda_2$ & $\lambda_3$ &  \\%
[0.1cm] \hline
&  &  &  &  \\
$E$ & $3(\omega_m+1)$ & $3(\omega_m+1)$ & $3(2\omega_m+1)/2$ &  \\[0.2cm]
$M$ & $-3(\omega_m+1)$ & $3(\omega_m+1)/2$ & $3\omega_m/2$ &  \\[0.2cm]
$K$ & $-3\omega_m$ & $-3$ & $3/2$ &  \\[0.2cm]
$MS$ & $-3\gamma_m$ & $\frac{3}{4}(1-\omega_m+\xi/\mu)$ & $-\frac{3}{4}%
(1-\omega_m-\xi/\mu)$ &  \\[0.2cm] \hline\hline
\end{tabular}%
\end{table*}

\section{Equilibrium Points in the Phase Space}

In this section we will analyze in detail the existence and stability of critical points of the autonomous systems corresponding to both
Randall-Sundrum and Dvali-Gabadadze-Porrari braneworld models. In both cases we study i) an AdS throat -- often explored in the literature --, and a
quadratic potential: $f(\phi)=\alpha/\phi^4$, and $V(\phi)=m^2\phi^2/2$ \cite{basic} and ii) a particular choice of the warp factor and of the
potential for the DBI field: $f(\phi)=1/V(\phi)$.

\subsection{An AdS throat and a quadratic potential}

Hence, the following relationship between $f$ and $V$:
\begin{equation*}
f=\frac{\alpha m^4}{4 V^2},
\end{equation*}
is obtained. The above choice leads to $\mu_{1}=\mu $ being a constant
\begin{equation*}
\mu =\sqrt{2/\alpha}(2/m),
\end{equation*}
while
\begin{equation*}
\mu_{2}=-2\gamma (x^{2}/z^{2})\mu.
\end{equation*}
Consequently, only two of the phase space variables $x$, $z$, and $\mu_{2}$,
are independent (say $x$ and $z$). This fact leads to considerable
simplification of the problem since the dimension of the autonomous system
reduces from six down to four. This is one of the reasons why the present
particular case is generously considered in the bibliography.\footnote{%
The importance of the quadratic potential in the cosmological context is
remarkable \cite{Ahn,Copeland2}.}

\subsubsection{DBI-RS Model} \label{RSa}

As just noticed, after considering the specific form of the functions $%
f(\phi)$ and $V(\phi)$ above, the six-dimensional autonomous dynamical
system (\ref{eqx}-\ref{eqm2}) can be reduced down to a four-dimensional one:

\begin{eqnarray}
&&x^{\prime}=-\frac{\mu yz}{2x^{2}}(z^{2}-2\gamma x^{2})-\frac{y^{2}}{2x}-x%
\frac{H^{\prime }}{H},  \notag \\
&&y^{\prime}=3\mu\gamma (1-\gamma)^{2} xz-\frac{3\mu (1+\gamma^{2})z^{3}}{2x}
\notag \\
&&\;\;\;\;\;\;\;\;\;\;\;\;\;\;\;\;\;\;\;\;\;\;\;\;-\frac{3(1+\gamma ^{2})y}{2%
}-y\frac{H^{\prime }}{H},  \label{asode1} \\
&&z^{\prime}=\frac{\mu yz^{2}}{2x}-z\frac{H^{\prime}}{H},\;\;r^{\prime}=%
\frac{2r(r-1)}{2-r}\frac{H^{\prime }}{H},  \notag
\end{eqnarray}
where the ratio $H^{\prime}/H=-(q+1)$ is given by Eq. (\ref{h'/h}). It
arises the following constraint:

\begin{equation}
\Omega_{m}=\frac{\rho_{m}}{3H^{2}}=r-(1-\gamma)x^{2}-z^{2}.
\label{constraint}
\end{equation}
Since $\Omega _{m}\geq 0$, then $(1-\gamma )x^{2}+z^{2}\leq r$. Besides,
since $\Omega _{m}\leq 1$ then $(1-\gamma )x^{2}+z^{2}\geq r-1$. We will be
focused on expanding FRW universes, so that $x\geq 0$ and $z\geq 0$. The
resulting four-dimensional phase space for the DBI-RS model is the following:

\begin{eqnarray}
&&\Psi=\{(x,y,z,r):r-1\leq (1-\gamma )x^{2}+z^{2}\leq r,  \notag \\
&&\;\;\;\;\;\;\;\;\;\;\;\;\;x\geq 0,\;y^{2}\leq 3x^{2},\;z\geq 0,\;0\leq
r\leq 1\}.  \label{phasespaceA}
\end{eqnarray}

As properly noticed in the former section equilibrium points liying on the
hyper-plane $\sum=(x,y,z,r=1)$ are associated with standard general
relativity dynamics. The remaining points in $\Psi$ are related with
higher-dimensional brane effects.

The relevant equilibrium points of the autonomous system of equations (\ref%
{asode1}) are summarized in table \ref{tab1}, while the eigenvalues of the
corresponding linearization (Jacobian) matrices are shown in table \ref{tab2}%
. The empty (Misner-RS) universe (critical point $E$ in table \ref{tab1}) is the past attractor of the RS
cosmological model and a decelerated solution whenever $\omega
_{m}>-2/3$ otherwise is an accelerated solution and is a saddle point. Existence of the
empty universe is a distinctive feature of the higher-dimensional (brane) contributions.

\begin{figure}[th]
\begin{center}
\includegraphics[width=4.1cm,height=4cm]{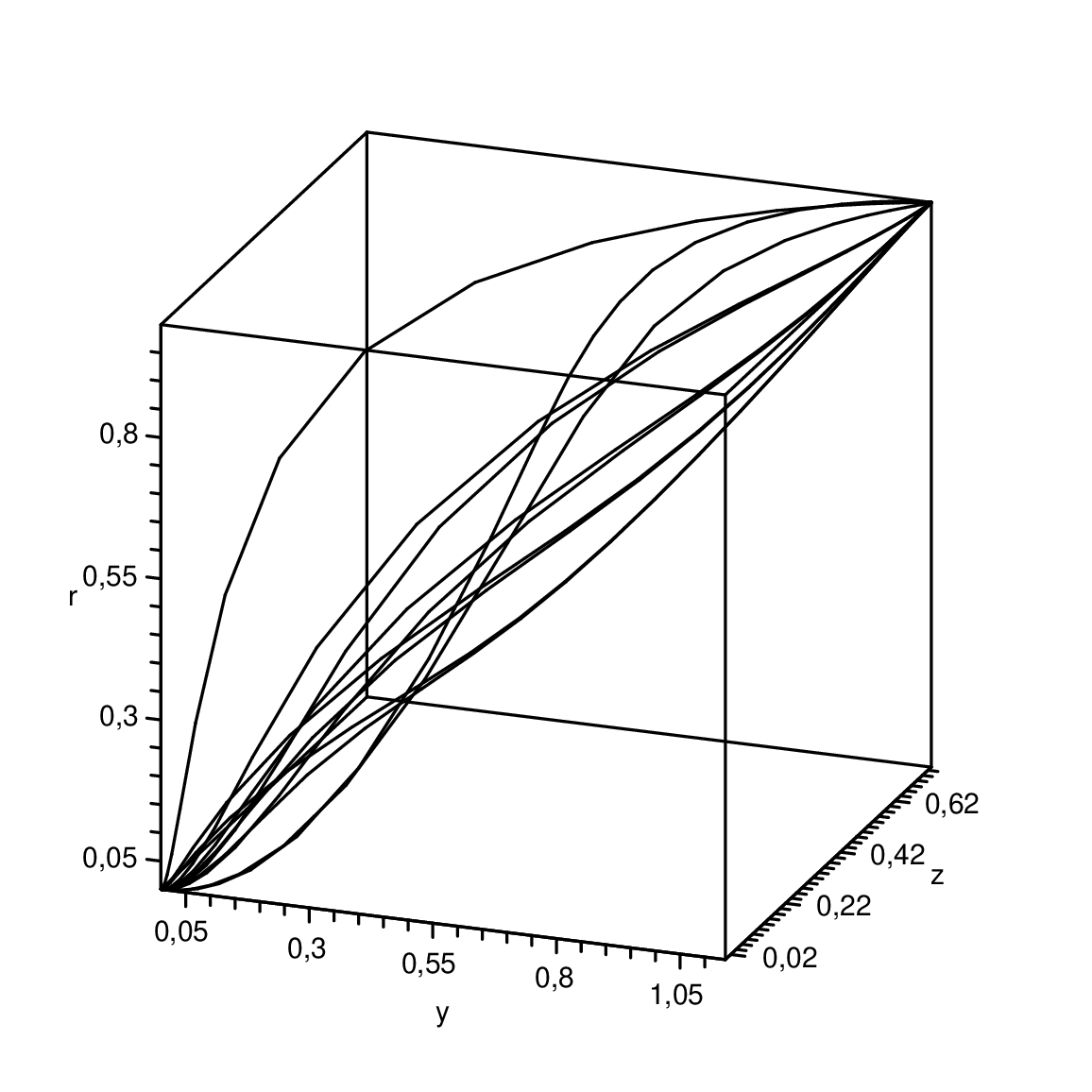}\vspace{0.3cm}%
\includegraphics[width=4.1cm,height=4cm]{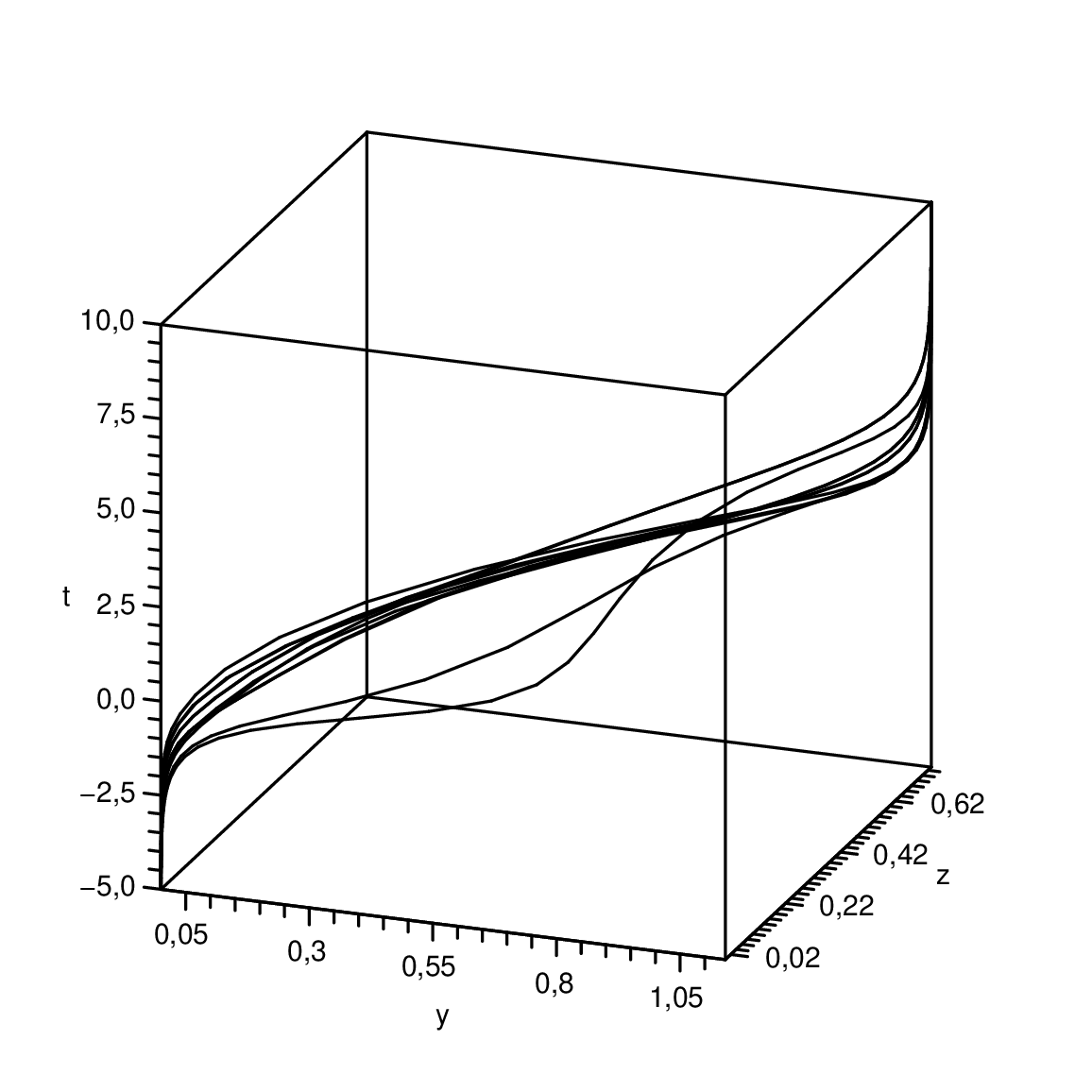}\vspace{0.3cm}\bigskip
\end{center}
\caption{Phase portrait generated by a given set of initial data ($\protect%
\omega _{m}=1/3$, $\protect\mu =1$) for the autonomous system of ODE (%
\protect\ref{ultra1}), corresponding to the UR approximation of the DBI-RS
model -- left-hand panel, and the corresponding flux in time -- right-hand
panel. The values of the free parameter has been chosen so that the scalar
field-dominated solution (point SF ($\pm 1.14,0.75,1$)) is the late-time
attractor.}
\label{fig1}
\end{figure}

The matter-dominated solution (equilibrium point $M$) and the
ultra-relativistic (scalar field-dominated) phase $U^{\pm }$, are always
saddle critical points and are associated with standard general relativity
dynamics. The latter is associated with decelerating dynamics while the
former $M$ represents decelerating expansion whenever $\omega _{m}>-1/3$.
Points $M$ and $U^{\pm }$ in Tab. \ref{tab1} correspond to the equilibrium
points $A$ and $B$ of Ref. \cite{basic}, respectively.

More detailed information can be retrieved only after further simplification
of the case to study. One way to achieve further simplification is to study
the ultra-relativistic regime where $\gamma =0\;\Rightarrow \;\gamma
_{L}=\infty $. In the UR regime, thanks to the relationship
\begin{equation*}
\gamma =0\;\;\Rightarrow \;\;x=\pm y/\sqrt{3},
\end{equation*}%
the autonomous system of ODE (\ref{asode1}) simplifies down to a set of
three ordinary differential equations:

\begin{eqnarray}
y^{\prime } &=&-\frac{3\sqrt{3}\mu z^{3}}{2y}-\frac{3y}{2}-y\frac{H^{\prime }%
}{H},  \notag \\
z^{\prime } &=&\frac{\sqrt{3}\mu z^{2}}{2}-z\frac{H^{\prime }}{H}%
,\;\;r^{\prime }=\frac{2r(r-1)}{2-r}\frac{H^{\prime }}{H},  \notag \\
\frac{H^{\prime }}{H} &=&-\frac{2-r}{2r}\{(\omega
_{m}+1)[3r-y^{2}-3z^{2}]+y^{2}\}.  \label{ultra1}
\end{eqnarray}

The reduced (three-dimensional) phase space in this simpler case is given by:

\begin{eqnarray}
\Psi &=&\{(y,z,r):3(r-1)\leq y^{2}+3z^{2}\leq 3r,  \notag \\
&&y\in \Re ,\;z\geq 0,\;0<r\leq 1\}.  \label{phasespaceA1}
\end{eqnarray}

There are found five equilibrium points of the autonomous system of ODE (\ref%
{ultra1}) in $\Psi $. These points, together with their properties, are
listed in Tab.\ref{tab1'}, while the eigenvalues of the corresponding
linearization matrices are shown in the table \ref{tab2'}. By their
overwhelming complexity, the eigenvalues of the linearization matrix
corresponding to the fourth equilibrium point in table \ref{tab1'} (point $%
SF $) have not been included in Tab.\ref{tab2'}. It is worth recalling that
the standard general relativity behaviour is associated with points in the
phase space liying on the phase plane $Hyp=(y,z,r=1)$.

The only one point that belongs to the bulk of the phase space $\Psi $ that is associated with 5D effects is again the point $E.$
Provided that $\omega _{m}>-1/2$ the empty universe --equilibrium point $E$
in Tab. \ref{tab1'}-- is always the past attractor in the phase space,
\textit{i.e.}, it represents the source critical point from which any phase
path in $\Psi $ originates and it is decelerated point.

The matter-scaling solution (equilibrium point $MS$) is the late-time
attractor provided that (the definition of the parameter $\xi $ can be found
in the caption of the table \ref{tab2'})
\begin{equation*}
-(1-\omega _{m})\mu <\xi <(1-\omega _{m})\mu .
\end{equation*}%
Otherwise, the scalar field-dominated solution $SF$ is the late-time
attractor.

We have to point out that the matter-scaling solution exists only if the
equation of state (EOS) parameter of the background matter is a negative
quantity: $-1<\omega _{m}<0$. This means that we can not have matter-scaling
with background matter being dust ($\omega _{m}=0$). Therefore, the
usefulness of this equilibrium point to describe the current phase of the
cosmic evolution is unclear. Unlike this, the scalar field-dominated solution $SF$ is always inflationary
($\eta >0$ always) and could be associated with accelerated late-time cosmic
dynamics. The matter-dominated solution $M$ and the kinetic energy-dominated
phase (equilibrium point $K$ in Tab. \ref{tab1'}), are always saddle points
in the phase space.

Worth noticing that the scalar field-dominated solution (point $SF$ in Tab.%
\ref{tab1'}) and the matter-scaling phase $MS$ correspond to the points $C$
and $D$ in reference \cite{basic}. In the standard $4$D limit of the theory we obtained the same behavior that in \cite{basic,Copeland2}.

In the Figure \ref{fig1} the trajectories in $\Psi $ -- the reduced phase
space defined in (\ref{phasespaceA1}) -- originated by a given set of
appropriated initial data, are drawn for the model of (\ref{asode1}) in the
ultra-relativistic approximation. The phase space pictures in the figures %
\ref{fig1} reveal the actual behavior of the RS dynamics: trajectories in
phase space depart from the (singular) empty universe, possibly related with
the initial (Big-Bang) singularity, and, at late times, approach to the plane ($y,z,1$), which is associated
with standard four-dimensional behavior, in particular the trajectories
approach to the $SF$ point.

\begin{table*}[tbp]
\caption[crit]{Properties of the equilibrium points of the autonomous system
(\protect\ref{asode2}).}%
\begin{tabular}{@{\hspace{4pt}}c@{\hspace{14pt}}c@{\hspace{14pt}}c@{\hspace{14pt}}c@{\hspace{14pt}}c@{\hspace{14pt}}c@{\hspace{14pt}}c@{\hspace{14pt}}c@{\hspace{14pt}}c@{\hspace{14pt}}c@{\hspace{14pt}}c}
\hline\hline
&  &  &  &  &  &  &  &  &  &  \\[-0.3cm]
Equilibrium point & $x$ & $y$ & $z$ & $r$ & Existence & $\gamma $ & $\Omega
_{m}$ & $\Omega _{\phi }$ & $\omega _{\phi }$ & $q$ \\ \hline
$E$ & $0$ & $0$ & $0$ & $0$ & Always & Undefined & $0$ & $0$ & Undefined & $%
-1$ \\
$M$ & $0$ & $0$ & $0$ & $1$ & " & " & $1$ & $0$ & " & $(1+3\omega _{m})/2$ \\%
[0.2cm]
$U^{\pm }$ & $1$ & $\pm \sqrt{3}$ & $0$ & $1$ & " & $0$ & $0$ & $1$ & $0$ & $%
1/2$ \\[0.2cm]\hline\hline
\end{tabular}%
\newline
\label{tab3}
\end{table*}

\begin{table*}[tbp]
\caption[eigenv]{Eigenvalues of the linearization matrices corresponding to
the critical points in table \protect\ref{tab3}.}%
\begin{tabular}{@{\hspace{4pt}}c@{\hspace{14pt}}c@{\hspace{14pt}}c@{\hspace{14pt}}c@{\hspace{14pt}}c}
\hline\hline
&  &  &  &  \\[-0.3cm]
Equilibrium point & $\lambda _{1}$ & $\lambda _{2}$ & $\lambda _{3}$ & $%
\lambda _{4}$ \\[0.1cm]\hline
$E$ & $3(1+\omega _{m})/2$ & $3(1+\omega _{m})/2$ & $-3(1+\omega _{m})/2$ & $%
-3(1-\omega _{m})/2$ \\
$M$ & $3(\omega _{m}-1)/2$ & $3(1+\omega _{m})/2$ & $3(1+\omega _{m})/2$ & $%
3(1+\omega _{m})/2$ \\[0.2cm]
$U^{\pm }$ & $-3\omega _{m}$ & $3$ & $3/2$ & $3/2$ \\[0.2cm]\hline\hline
\end{tabular}%
\label{tab4}
\end{table*}

\begin{table*}[tbp]
\caption[crit]{Properties of the equilibrium points of the autonomous system
(\protect\ref{ultra2}). Here $\protect\eta \equiv \protect\mu (\protect\sqrt{%
\protect\mu ^{2}+12}-\protect\mu )$, while $\protect\gamma _{m}\equiv
\protect\omega _{m}+1$.}
\label{tab3'}%
\begin{tabular}{@{\hspace{4pt}}c@{\hspace{14pt}}c@{\hspace{14pt}}c@{\hspace{14pt}}c@{\hspace{14pt}}c@{\hspace{14pt}}c@{\hspace{14pt}}c@{\hspace{14pt}}c@{\hspace{14pt}}c}
\hline\hline
&  &  &  &  &  &  &  &  \\[-0.3cm]
Equilibrium point & $y$ & $z$ & $r$ & Existence & $\Omega _{m}$ & $\Omega
_{\phi }$ & $\omega _{\phi }$ & $q$ \\ \hline
&  &  &  &  &  &  &  &  \\
$M$ & $0$ & $0$ & $1$ & " & $1$ & $0$ & Undefined & $\frac{3\omega _{m}+1}{2}
$ \\[0.2cm]
$K$ & $\pm \sqrt{3}$ & $0$ & $1$ & " & $0$ & $1$ & $0$ & $\frac{1}{2}$ \\%
[0.2cm]
$SF$ & $\pm \sqrt{\frac{\eta }{2}}$ & $\frac{\eta }{(2\sqrt{3}\mu )}$ & $1$
& $\mu \geq 0$ & $0$ & $1$ & $-1+\frac{\eta }{6}$ & $-1-\frac{\eta }{4}$ \\%
[0.2cm]
$MS$ & $\pm \frac{3\gamma _{m}}{\mu }\sqrt{-\frac{\gamma _{m}}{\omega _{m}}}$
& $\frac{\sqrt{3}\gamma _{m}}{\mu }$ & $1$ & $-1<\omega _{m}<0$ & $1+\frac{%
3\gamma _{m}^{2}}{\mu ^{2}\omega _{m}}$ & $-\frac{3\gamma _{m}^{2}}{\mu
^{2}\omega _{m}}$ & $\omega _{m}$ & $\frac{3\omega _{m}+1}{2}$ \\%
[0.2cm] \hline\hline
\end{tabular}
\newline
\end{table*}

\begin{table*}[tbp]
\caption[eigenv]{Eigenvalues of the linearization matrices corresponding to
the first four critical points in table \protect\ref{tab3'}. Here $\Pi\equiv%
\protect\sqrt{12\protect\omega_m(6\protect\omega_m+\protect\mu\protect\eta)+%
\protect\mu^{2}(6-\protect\mu\protect\eta)}$, while $\protect\zeta\equiv%
\protect\sqrt{24(1+\protect\omega _{m})^{3}+\protect\mu ^{2}(1+3\protect%
\omega _{m})^{2}.}$}
\label{tab4'}%
\begin{tabular}{@{\hspace{4pt}}c@{\hspace{14pt}}c@{\hspace{14pt}}c@{\hspace{14pt}}c@{\hspace{14pt}}c}
\hline\hline
&  &  &  &  \\[-0.3cm]
Equilibrium point & $\lambda_1$ & $\lambda_2$ & $\lambda_3$ &  \\%
[0.1cm] \hline
&  &  &  &  \\
$M$ & $3(\omega_m+1)/2$ & $3(\omega_m+1)/2$ & $3\omega_m/2$ &  \\[0.2cm]
$K$ & $-3\omega_m$ & $3/2$ & $3/2$ &  \\[0.2cm]
$SF$ & $\mu \eta/4$ & $\frac{-12(2+\omega_{m})+3\mu\eta}{8}+\frac{\sqrt{2}}{8%
}\Pi$ & $\frac{-12(2+\omega_{m})+3\mu\eta}{8}-\frac{\sqrt{2}}{8}\Pi$ &  \\%
[0.2cm]
$MS$ & $\frac{3}{2}(1+\omega _{m})$ & $\frac{3}{4}\left( \omega _{m}-1+\frac{%
\zeta }{\mu^{2}}\right) $ & $\frac{3}{4}\left( \omega _{m}-1-\frac{\zeta }{%
\mu^{2}}\right) $ &  \\[0.2cm] \hline\hline
\end{tabular}%
\end{table*}

\subsubsection{DBI-DGP Model} \label{DGPa}

For an AdS throat and the quadratic self-interaction potential, the
autonomous system of ODE (\ref{eqx'})-(\ref{eqzr'}), (\ref{eqm1}), (\ref%
{eqm2}), reduces down to the following four-dimensional autonomous system:

\begin{eqnarray}
x^{\prime } &=&-\frac{y^{2}}{2x}-\frac{\mu yzr}{2x^{2}}(z^{2}-2\gamma
x^{2})-x\frac{Q^{\prime }}{Q},  \notag \\
y^{\prime } &=&-6\gamma ^{2}\mu xzr-\frac{3(\gamma ^{2}+1)}{2x}[\mu
(z^{2}-2\gamma x^{2})zr+  \notag \\
&&+xy]-y\frac{Q^{\prime }}{Q},  \notag \\
z^{\prime } &=&\frac{\mu yz^{2}r}{2x}-z\frac{Q^{\prime }}{Q},\;\;r^{\prime
}=r\left( \frac{1-r^{2}}{1+r^{2}}\right) \frac{Q^{\prime }}{Q}.
\label{asode2}
\end{eqnarray}%
The ratio $Q^{\prime }/Q$ is given by Eq. (\ref{q'/q}).

Recall that standard general relativity behavior is associated with points
liying on the hyper-plane $Hyp=(x,y,z,r=1)$. The remaining points in the
bulk of the phase space $\Psi_\pm$ correspond to higher-dimensional
behaviour.

The critical points of the autonomous system of ODE (\ref{asode2}), together
with their most important properties, are summarized in the table \ref{tab3}%
. The eigenvalues of the linearization matrices corresponding to the
critical points in Tab. \ref{tab3} are shown in table \ref{tab4}.

As for the DBI field trapped in a RS brane, in the present case the
matter-dominated solution ($M$), the empty universe ($E$) and the UR phase ($U^{\pm }$) -- dominated
by the scalar field --, are found. The critical point $E$ (the empty (Misner-DGP) universe) is always a saddle point and an accelerated solution. The existence of this point is an unexpected results. We can associated this point (when $r=0$) to two possibility: a) to early time, when $H\rightarrow\infty$,  or b) late time, when $Q=0$ and $H=\pm1/r_c$\footnote{%
In fact, fitting SN observations requires $H\geq1/r_c$ in order to achieve late time acceleration (see, for instance, reference \cite{Koyama} and references therein).}. The matter-dominated solution (critical point $M$ in Tab. \ref{tab3}) have a same behavior than the one found in the Randall-Sundrum case. The ultra-relativistic regime that is dominated by the
scalar field mimics the cosmic evolution of a universe filled with dust. The above
results are to be contracted with the results in the former subsection.

As before, a more detailed study of the asymptotic properties of the model (%
\ref{asode2}) requires additional simplification. The ultra-relativistic
approximation comes to our rescue. As long as one considers just large
Lorentz boosts (amounting to vanishing $\gamma $) the relationship $x=\pm
y/\sqrt{3}$ is verified. This relationship allows for further simplification
of the autonomus system of ODE (\ref{asode2}). Actually, in the UR regime
the above system of equations can be simplified to the following
three-dimensional autonomous system of ODE:

\begin{eqnarray}
y^{\prime } &=&-\frac{3}{2}y-\frac{3\sqrt{3}\mu z^{3}r}{2y}-y\frac{Q^{\prime
}}{Q},  \notag \\
z^{\prime } &=&\frac{\sqrt{3}}{2}\mu z^{2}r-z\frac{Q^{\prime }}{Q},  \notag
\\
r^{\prime } &=&\frac{r(1-r^{2})}{(1+r^{2})}\frac{Q^{\prime }}{Q}.
\label{ultra2}
\end{eqnarray}%
The phase space for the autonomous system (\ref{ultra2}) can be defined in
the following way. For the "+" - branch (the Minkowski cosmological phase):

\begin{eqnarray}
\Psi _{+} &=&\{(y,z,r):\;0\leq y^{2}+3z^{2}\leq 3,  \notag \\
&&y\in \Re ,\;\;z\geq 0,\;\;r\in \lbrack 1,\infty )\},  \label{PhaseSpace+}
\end{eqnarray}%
while, for the self-accelerating "-" - branch, it is given by:

\begin{eqnarray}
\Psi _{-} &=&\{(y,z,r):\;0\leq y^{2}+3z^{2}\leq 3,  \notag \\
&&y\in \Re ,\;\;z\geq 0,\;\;r\in ]0,1]\}.  \label{PhaseSpace-}
\end{eqnarray}

There are four equilibrium points of the autonomous system of ODE(\ref%
{ultra2}). These critical points -- together with their most salient
features -- are summarized in table \ref{tab3'}.

If $\omega _{m}>0$ them the matter-dominated solution (point $M$) is the
past attractor else the kinetic energy-dominated solution (point $K$) is the
past attractor in the phase space, as in \cite{basic,Copeland2}, independent on which
branch of the DGP is being considered since, at early times, the brane
effects can be safely ignored so that the standard cosmological dynamics is
not modified.

The equilibrium point $SF$ (scalar field-dominated solution) and $MS$
(scaling dominated-solution) are always a saddle point. These solutions
show a quite different behaviour than the one found in the Randall-Sundrum
case. There is no future (late-time) attractor in the
phase space of the model .

\subsection{When $f(\phi)=1/V(\phi)$}

This case was studied in \cite{deformedtaquion} where the author presented approach relies on an existing (formal) mathematical equivalence between a Dirac-Born-Infeld (DBI) model and standard tachyon cosmology, under an appropriate transformation of the DBI field. The above choice leads to significant simplification of the field equations.

The equation of motion of the DBI-type field and the conservation equation are the following:

\be \ddot\phi+3\gamma^{-2}H\dot\phi=-\partial_\phi
V(1-3\dot\phi^2/2V),\label{feqsmod}\ee where the modified Lorentz
factor $\gamma$ is defined as:
\be\gamma=\frac{1}{\sqrt{1-\dot\phi^2/V}},
\label{modlorentzfactor}\ee and where we have defined the following energy density and pressure of the DBI scalar field:
\be \rho_\phi=\gamma V(\phi)\;,\;\;p_\phi=-\gamma^{-1}V(\phi).\nonumber\ee

From now on we shall call the model given by equations
(\ref{feqsmod}), (\ref{modlorentzfactor}), as modified tachyon
cosmology (MTC).

Whit this choice we reduce the autonomous system to a three
dimension because we can introduce a new dimensionless variable
$a\equiv y/x$ and we have that $\mu_1=\partial_\phi V/V=-\mu_2$.

\subsubsection{DBI-RS Model} \label{RSb}

With the aim to obtained an autonomous system  of the cosmological equations we introduce the following dimensionless phase space variables: \be
a=\frac{\dot\phi}{\sqrt{V}},\;z=\frac{\sqrt{V}}{\sqrt{3}H},\;
r=\frac{\rho_T}{3H^2}.\ee We can realize that

\bea \frac{\rho_T}{\lambda}=\frac{2(1-r)}{r},\;\Rightarrow\;0<r\leq
1.\label{zConstraint}\eea

Then we can write the above system as: \bea
&&a'=(a^2-1)[3a+\sqrt{3}\;\partial_\phi(\ln V)z]\nonumber\\
&&z'=\sqrt{\frac{3}{4}}az^2\;\partial_\phi(\ln V)-\nonumber\\
&&-\frac{z(2-r)}{2r}\left[\frac{3z^2(\gamma_{m}-a^2)}{\sqrt{1-a^2}}-3\gamma_m r\right]\nonumber\\
&&r'=(r-1)\left[\frac{3z^2(\gamma_{m}-a^2)}{\sqrt{1-a^2}}-3\gamma_{m}r\right]\label{Systemod}\eea where $\gamma_{m}$ is the barotropic index for
the matter ($0\leq\gamma_{m}\leq 2$), the comma represent derivative
respect to cosmological time $\tau.$

The constrain of the Friedmann equatin is rewrite in the new
variables as: \bea \Omega_{m}=r-\frac{z^2}{\sqrt{1-a^2}}.\eea

The others cosmological parameters such as the energy density
parameter of the scalar field $\Omega_{\phi},$ the equation state
parameter $\omega_{\phi}$ and the deceleration parameter $q$ in the
new variables have the followings forms:
\bea
\Omega_{\phi}=\frac{z^2}{\sqrt{1-a^2}},\omega_{\phi}=a^2-1\nonumber\\
q=-1+\frac{(2-r)}{2r}\left[-3\gamma_{m}r+3\gamma
z^2(\gamma_{m}-a^2)\right]\eea

Finally the phase space for this variables is

\bea \Psi=\{(a,z,r):-1\leq a\leq 1,0\leq z^4\leq 1-a^2,\nonumber\\
0<r\leq1\} \label{psimod}\eea

For an exponential self-interaction potential of the form: \be
V(\phi)=V_0 \exp (-\lambda\phi),\nonumber\ee since $\partial_\phi\ln
V=-\lambda=const$, then the system ($\ref{Systemod}$) is a closed
autonomous system: \bea
&&a'=(a^2-1)[3a-\sqrt{3}\lambda z]\nonumber\\
&&z'=-\sqrt{\frac{3}{4}}\lambda az^2-\frac{z(2-r)}{2r}\left[\frac{3z^2(\gamma_{m}-a^2)}{\sqrt{1-a^2}}-3\gamma_m r\right]\nonumber\\
&&r'=(r-1)\left[\frac{3z^2(\gamma_{m}-a^2)}{\sqrt{1-a^2}}-3\gamma_{m}r\right]\label{Systemod1}\eea

\begin{table*}[tbp]\caption[crit]{Properties of the critical points for the autonomous system (\ref{Systemod}). Where $z_*=\sqrt{\frac{\sqrt{36+\lambda^4}-\lambda^2}{6}}$.}
\begin{tabular}{@{\hspace{4pt}}c@{\hspace{14pt}}c@{\hspace{14pt}}c@{\hspace{14pt}}c@{\hspace{14pt}}c@{\hspace{14pt}}c@{\hspace{14pt}}c@{\hspace{14pt}}c}
\hline\hline\\[-0.3cm]
Equilibrium Point &$a$&$z$&$r$&Existence& $\Omega_\phi$& $\omega_\phi$ & $q$\\[0.1cm]
\hline\\[-0.2cm]
%%%%%%%%
$M$ & $0$ & $0$ &$1$&  Always & $0$ & $-1$ & $-1+\frac{3\gamma_{m}}{2}$ \\[0.2cm]
$U^{\pm}$ & $\pm1$ & $0$ & $1$ & '' & $0$ & $0$ & $-1+\frac{3\gamma_{m}}{2}$ \\[0.2cm]
$MS$ & $\sqrt{\gamma_m}$ & $\frac{\sqrt{3\gamma_m}}{\lambda}$ &$1$& ''&$\frac{3\gamma_m}{\sqrt{1-\lambda^2\gamma_m}}$&$\omega_m$&$-1+\frac{3\gamma_m}{2}$ \\[0.2cm]
$T$&$\frac{\lambda z_*}{\sqrt{3}}$&$z_*$&$1$& ''&$1$&$-1-\frac{\lambda^2z_*^2}{3}$&$-1+3\gamma_{m}/2$\\[0.2cm]
\hline \hline
\end{tabular}\label{tabmodCaso1}
\end{table*}

\begin{table*}[tbp]\caption[eigenv]{Eigenvalues of the linearization matrice corresponding to the equilibrium points in Tab.\ref{tabmodCaso1}.\\Here $\Pi\equiv\sqrt{\frac{48\gamma_m^2\sqrt{1-\gamma_m}}{\lambda^2}+4+\gamma_m(17\gamma_m-20)}$.}
\begin{tabular}{@{\hspace{4pt}}c@{\hspace{14pt}}c@{\hspace{14pt}}c@{\hspace{14pt}}c}
\hline\hline\\[-0.3cm]
Equilibrium Point& $\lambda_1$& $\lambda_2$& $\lambda_3$\\[0.1cm]\hline\\[-0.2cm]
%%%%%%%%
$M$ & $-3$ &$-3\gamma_{m}$& $3\gamma_{m}/2$\\[0.2cm]
$U^{\pm}$ & $-3\gamma_{m}$ &$3\gamma_{m}/2$& $6$\\[0.2cm]
$MS$ & $-3\gamma_m$ &$-\frac{3}{4}\left[(2-\gamma_m)+\Pi\right]$& $-\frac{3}{4}\left[(2-\gamma_m)-\Pi\right]$\\[0.2cm]
$T$ & $-\frac{\lambda^2\left(\sqrt{\lambda^4+36}-\lambda^2\right)}{12}$ & $-3+\frac{\lambda^2}{12}\left(\sqrt{\lambda^4+36}-\lambda^2\right)$ & $-3\gamma_m+\frac{\lambda^2}{12}\left(\sqrt{\lambda^4+36}-\lambda^2\right)$\\[0.2cm]
\hline \hline
\end{tabular}\label{tabmodCaso11}
\end{table*}
\begin{figure}[th]
\begin{center}
\includegraphics[width=4.1cm,height=4cm]{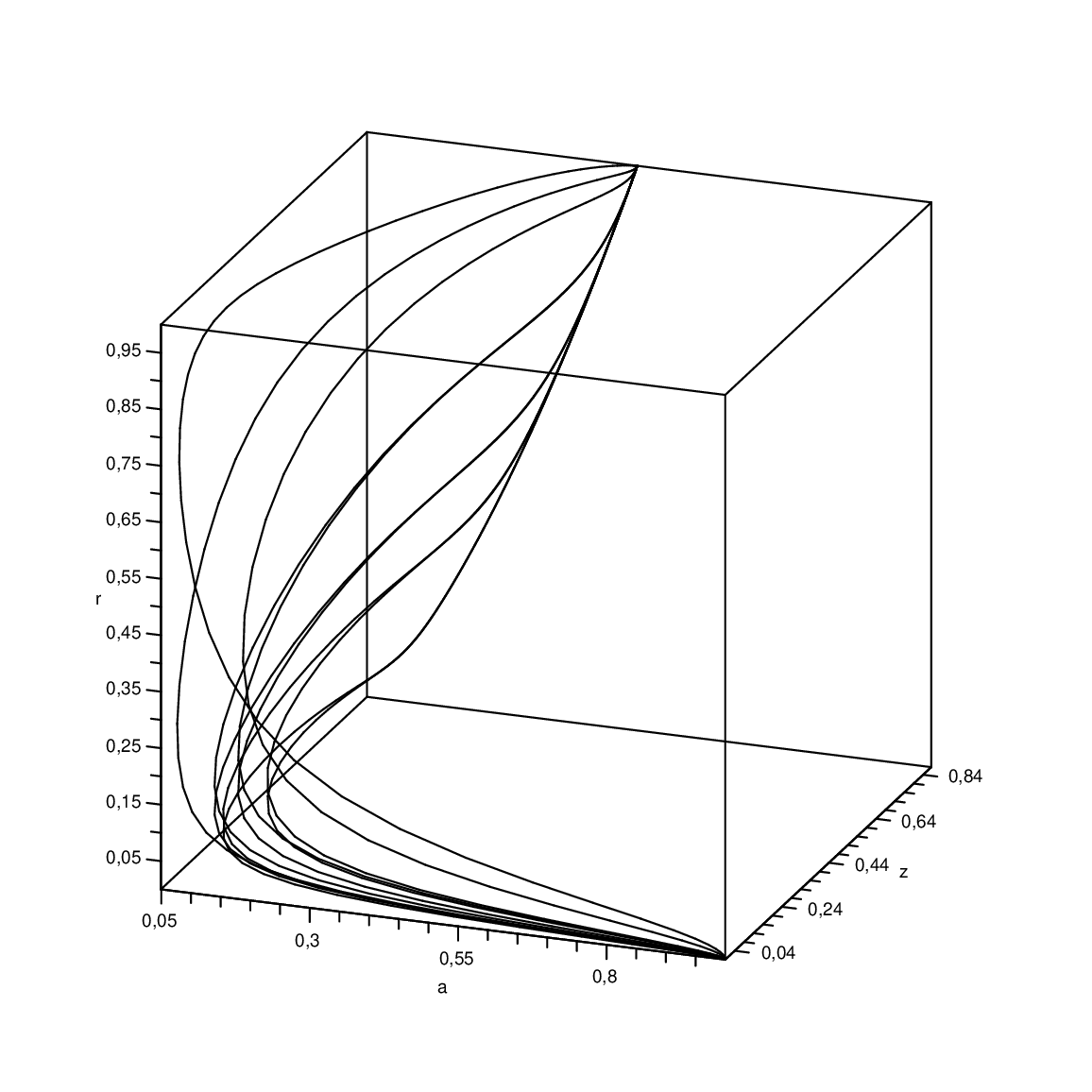}\vspace{0.3cm}%
\includegraphics[width=4.1cm,height=4cm]{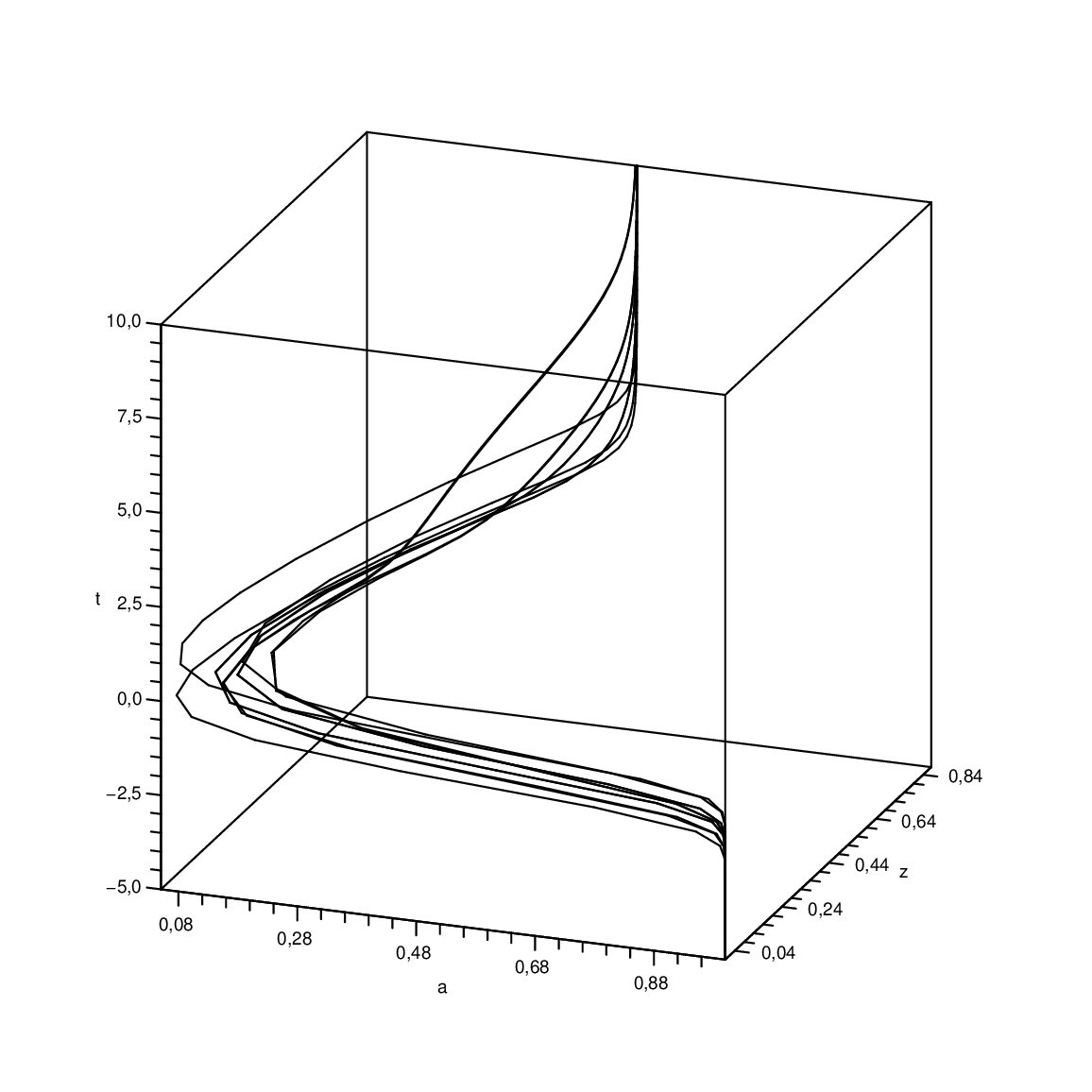}\vspace{0.3cm}\bigskip
\end{center}
\caption{Phase portrait generated by a given set of initial data ($\protect%
\omega _{m}=1$, $\protect\lambda =1$) for the autonomous system of ODE (%
\protect\ref{Systemod1}), corresponding to the UR approximation of
the DBI-RS model -- left-hand panel, and the corresponding flux in
time -- right-hand panel. The trajectories in phase space emerge
from the point $S=(a,z,r)=(\pm1,0,0)$- the empty Misner-RS
universe.} \label{figtaqrs}
\end{figure}

The critical points of this potential are summarized in Table
\ref{tabmodCaso1} and it´s respective eigenvalues in Table
\ref{tabmodCaso11}. All this point are the same that was obtained in
\cite{quiros} and the point $M$, $T$ and $MS$ have a similar
stability property. The stability of point $U$ differ from the yours
property in the standard cosmology \cite{quiros} where was a
inflationary past attractor and in this case is a saddle point.

In Figure \ref{figtaqrs} we show the trajectories in phase space for
different sets of initial conditions. As clearly seen from this
figure, the trajectories in phase space emerge from the point
$S=(a,z,r)=(\pm1,0,0)$- the empty Misner-RS universe- meaning that
this is the past attractor of the Randall-Sundrum cosmological
model. We want to notice that the points with $r=0$ have been
removed from the phase space $\Psi$ since, in general, at up due to
our choice of space variables. For this reason this point no appear
in Table \ref{tabmodCaso1}.

\begin{table*}[tbp]\caption[eigenv]{Eigenvalues of the linearization matrice corresponding to the equilibrium points $U$, $T$ and $MS$ to the system \ref{Systemod2}.\\Here $\Pi\equiv\sqrt{\frac{48\gamma_m^2\sqrt{1-\gamma_m}}{\lambda^2}+4+\gamma_m(17\gamma_m-20)}$.}
\begin{tabular}{@{\hspace{4pt}}c@{\hspace{14pt}}c@{\hspace{14pt}}c@{\hspace{14pt}}c}
\hline\hline\\[-0.3cm]
Equilibrium Point& $\lambda_1$& $\lambda_2$& $\lambda_3$\\[0.1cm]\hline\\[-0.2cm]
%%%%%%%%
$U^{\pm}$ & $3\gamma_{m}/2$ &$3\gamma_{m}/2$& $6$\\[0.2cm]
$MS$ & $\frac{3\gamma_m}{2}$ &$-\frac{3}{4}\left[(2-\gamma_m)+\Pi\right]$& $-\frac{3}{4}\left[(2-\gamma_m)-\Pi\right]$\\[0.2cm]
$T$ & $\frac{\lambda^2\left(\sqrt{\lambda^4+36}-\lambda^2\right)}{12}$ & $-3+\frac{\lambda^2}{12}\left(\sqrt{\lambda^4+36}-\lambda^2\right)$ & $-3\gamma_m+\frac{\lambda^2}{12}\left(\sqrt{\lambda^4+36}-\lambda^2\right)$\\[0.2cm]
\hline \hline
\end{tabular}\label{tabmodCaso22}
\end{table*}
\subsubsection{DBI-DGP Model} \label{DGPb}

In this model we use the same  $a$ and $z$ variables to the former
subsection and redefine the $r$ variable how $r=Q/H$.

With these variable we can write the following system: Then we can
\bea
&&a'=(a^2-1)[3a+\sqrt{3}\;\partial_\phi(\ln V)z]\nonumber\\
&&z'=z\left[\sqrt{\frac{3}{4}}az\;\partial_\phi(\ln V)-\frac{2r^2}{r^2-1}\frac{Q'}{Q}\right]\nonumber\\
&&r'=\frac{r(1-r^2)}{r^2+1}\frac{Q'}{Q}\label{Systemod2}\eea where
$\gamma_{m}$ is the barotropic index for the matter
($0\leq\gamma_{m}\leq 2$), the comma represent derivative respect to
cosmological time $\tau$ and \be
\frac{Q'}{Q}=-\frac{3}{2r^2}\left[\gamma_mr^2+\frac{z^2(a^2-\gamma_m)}{\sqrt{1-a^2}}\right]
\nonumber \ee

The density parameter of the scalar field $\Omega_\phi$ and the
equation state parameter $\omega_\phi$ have the same expression that
the former subsection. The deceleration parameter $q$ in the new
variables have the following forms: \be
q=-1-\frac{2r^2}{r^2+1}\frac{Q'}{Q}.\ee

The phase space for the autonomous system (\ref{Systemod2}) can be
defined in the following way. For the "+" - branch (the Minkowski
cosmological phase): \bea
\Psi _{+} &=&\{(a,z,r):\;-1\leq a\leq1,\;0\leq z^4+a^2\leq 1,  \notag \\
&&\;\;z\geq 0,\;\;r\in \lbrack 1,\infty )\}, \label{PhaseSpace+}
\eea%
while, for the self-accelerating "-" - branch, it is given by:

\begin{eqnarray}
\Psi _{-} &=&\{(a,z,r):\;-1\leq a\leq1,\;0\leq z^4+a^2\leq 1,  \notag \\
&&y\in \Re ,\;\;z\geq 0,\;\;r\in ]0,1]\}.  \label{PhaseSpace-}
\end{eqnarray}

The system (\ref{Systemod2}) is not closed autonomous system and for
this reason we use the exponential potential
($V=V_0\exp(-\lambda\phi)$) because to $\partial_\phi V/V=-\lambda$.

We obtained the same four critical point that was obtained in the
former subsection in the Randall-Sundrum Model. But in this case the
stability property of critical point $U$, $T$ and $MS$ are different
and the critical point $M$ have a similar stability property. The
eigenvalues of critical point $U$, $T$ and $MS$ we show in the table
\ref{tabmodCaso22}.

The equilibrium point $U=(\pm1,0,1)$ represent an inflationary
solution - past attractor - in the phase space. These point are
associated with ultra-relativistic behavior since $a=\pm1$. This
point represents a scaling of the potential and of the kinetic
energy of the scalar field. These result are the same that was
obtained in \cite{quiros}.

The matter scaling solution is always a saddle point these result is
differ that was obtained in \cite{quiros} were this point whenever
exists it was a stable equilibrium point. The equilibrium point $T$
(the tachyon-dominated solution) is the late time attractor.

\section{Results and discussion}

The importance of the brane effects for the cosmic dynamics is well known.
These effects can modify the general relativity laws at early times (UV
modifications), as well as at late times (IR modifications). Actually, while
the Randall-Sundrum brane model produces UV modifications of general
relativity, in the Dvali-Gabadadze-Porrati braneworld IR modifications of
the laws of gravity arise instead. In a similar way, the introduction of a
non-linear Dirac-Born-Infeld type of field might modify the cosmic dynamics
at early as well as at late times. This makes even more interesting the
study of the combined effect of a DBI-type field trapped in the braneworld.
Aim of the present paper has been, precisely, the study of the asymptotic
properties of the latter kind of cosmological models.

\subsection{RS Model}

The main result of Sec.\ref{RSa} and Sec.\ref{RSb} can be summarized as follows:

\begin{itemize}
  \item The empty (Misner-RS) universe (critical point $E$ in Tab.\ref{tab1}) is always the past attractor in the phase space, it represents the source critical point from which any phase path in $\Psi$ originates and a decelerated solution.
  \item In the Sec.\ref{RSa} the matter dominated solution (critical point $M$ in Tab.\ref{tab1}) and the ultra-relativistic (scalar field-dominated) phase (critical point $U\pm$) are always saddle critical point and are associated with standard general relativity dynamics.
  \item The late-time behavior in Sec.\ref{RSa} are associated with the critical point $MS$ (matter-scaling solution) and $SF$ (scalar field solution). The point $MS$ is the late-time attractor provided that (the definition of the parameter $\xi $ can be found in the caption of the table \ref{tab2'}) $-(1-\omega _{m})\mu <\xi <(1-\omega _{m})\mu$. Otherwise, the scalar field-dominated solution $SF$ is the late-time attractor. This result are the same that was obtained in the standard $4$D limit of the theory in \cite{basic,Copeland2}.
  \item In the Sec.\ref{RSb} the critical point $M$, $T$ and $MS$ have a similar stability property that in the standard cosmology \cite{quiros}. The stability of point $U$ differ from the yours property in the standard cosmology \cite{quiros} where was a inflationary past attractor and in this case is a saddle point.
\end{itemize}

In general, the dynamical behavior of the Randall-Sundrum model differs from the standard behavior within four-dimensional Einstein-Hilbert gravity only at early times (high-energy regime). The existence the empty (Misner-RS) universe to be contrasted with the standard four-dimensional result were the fluid-dominated solution or kinetic-dominated solution can be the past attractor \cite{basic,Copeland2,quiros}. The late-time cosmological dynamics, on the contrary, is not affected by the RS brane effects in any essential way.

\subsection{DGP Model}

From the analysis in Sec.\ref{DGPa} and Sec.\ref{DGPb}, the following important results can be summarized:
\begin{itemize}
  \item The matter-dominated solution (critical point $M$) always is a saddle point and a decelerated solution.
  \item In the Sec.\ref{DGPa} we found the empty (Misner-DGP) universe (critical point $E$ in Tab.\ref{tab3}). The existence of this critical point is an unexpected results. This solution is always a saddle critical point and an accelerated solution. This solution can be associated with intermediate-time in the cosmic evolution.
  \item The early time behavior in the Sec.\ref{DGPa} are associated with the matter-dominated solution or kinetic-dominated solution as in \cite{basic,Copeland2}, independent on which branch of the DGP is being considered since, at early times, the brane effects can be safely ignored so that the standard cosmological dynamics is not modified.
  \item There is no future (late-time) attractor in the phase space of the model in Sec.\ref{DGPa}. The scalar field-dominated solution (equilibrium point $SF$) and the scaling dominated-solution (equilibrium point $MS$) are always a saddle point.
  \item In the Sec.\ref{DGPb} we obtained the same four critical point that was obtained in the Randall-Sundrum model (Sec.\ref{RSb}). But in this case the stability property of critical point $U$, $T$ and $MS$ are different, independent on which branch of the DGP is being considered since, at late times, the brane effects can not be safely ignored so that the standard cosmological dynamics is modified.
  \item The critical point $U$ in Sec.\ref{DGPb} represent an inflationary solution - past attractor - in the phase space. This point represents a scaling of the potential and of the kinetic energy of the scalar field. This point are the same stability property that was obtained in \cite{quiros}.
\end{itemize}

However the DGP brane effects indeed modify the late-time cosmological dynamics through changing the stability of the corresponding (late-time) critical points. Actually, in the present case the scalar field-dominated solution as well as the scaling dominated-solution are always a saddle point. This result has to be confronted with the classical general relativity result where the above-mentioned solutions can be late-time attractors.

In the DBI-DGP case, at early times, the dynamics is general relativistic so that the stability properties of the matter-dominated phase in tables \ref{tab1'} and \ref{tab2'} just coincide with the results of \cite{basic,Copeland2}.

\section{Conclusion}

In the present paper we aimed at studying the asymptotic properties of a DBI-type field trapped in the braneworld -- for two case i) an AdS throat -- often explored in the literature --, and a quadratic potential: $f(\phi)=\alpha/\phi^4$, and $V(\phi)=m^2\phi^2/2$ \cite{basic} and ii) a particular choice of the warp factor and of the potential for the DBI field: $f(\phi)=1/V(\phi)$ --, by means of the dynamical systems tools. The
combined effect of the non-linear nature of the DBI field and of the higher-dimensional brane effects seem to produce a rich dynamics. Both brane
contributions and non-linear DBI effects can modify the general relativity laws of gravity at late times as well as at early times. Here we focused in
Randall-Sundrum and in Dvali-Gabadadze-Porrati braneworlds exclusively. In a sense this work can be considered as a natural completion of the Ref. \cite{basic,Copeland2} to consider the combined effect of the DBI-type field and of the braneworld.

We performed a thorough study of the phase space corresponding to the two scenarios. It is revealed that the empty universe, the matter-dominated
solution, and the ultra-relativistic phase, are common to both of them. However, the stability properties of these points differ from one scenario
to another. While in the DBI-RS model the empty (not necessarily inflationary) universe is always the past attractor, for the DBI-DGP scenario the past attractor is the matter-dominated solution or the kinetic dominated-solution. The interchange of the stability properties of the equilibrium points $E$ and $M$ or $K$ in the DBI-RS and in the DBI-DGP models can be easily explained as due to the impact UV modifications produced by the Randall-Sundrum brane, have on the early-time cosmic dynamics.

This work was partly supported by CONACyT M\'{e}xico, under grants 49924-J, 52327, 105079, Instituto Avanzado de Cosmologia (IAC) collaboration. R G-S acknowledges partial support from COFAA-IPN, EDI-IPN and IPN grant SIP-20100610. D G, T G and I Q aknowledge also the MES of Cuba for partial
support of the research.


\begin{thebibliography}{99}
\bibitem{bennet} C. L. Bennett et al., Astrophys. J. Suppl. Ser. \textbf{148}
(2003) 1; G. Hinshaw et al., Astrophys. J. Suppl. Ser. \textbf{148} (2003)
135; D. N. Spergel et al., Astrophys. J. Suppl. Ser. \textbf{148} (2003)
175; H. V. Peiris et al., Astrophys. J. Suppl. Ser. \textbf{148} (2003) 213;
A. Kogut et al., Astrophys. J. Suppl. Ser. \textbf{148} (2003) 161; E.
Komatsu et al., Astrophys. J. Suppl. Ser. \textbf{148} (2003) 119

\bibitem{starobinsky} A. A. Starobinsky, Phys. Lett. B \textbf{91} (1980)
99; A. H. Guth, Phys. Rev. D \textbf{23} (1981) 347; A. Albrecht, P. J.
Steinhardt, Phys. Rev. Lett. \textbf{48} (1982) 1220; A. D. Linde, Phys.
Lett. B \textbf{108} (1982) 389; Phys. Lett. B \textbf{129} (1983) 177.

\bibitem{rs} L. Randall, R. Sundrum, Phys. Rev. Lett. \textbf{83} (1999)
3370.

\bibitem{dgp} G. R. Dvali, G. Gabadadze, M. Porrati,
%"4-D gravity on a brane in 5-D Minkowski space",
Phys. Lett. B\textbf{485} 208-214 (2000), [hep-th/0005016].

\bibitem{hawkins} R. M. Hawkins, J. E. Lidsey, Phys. Rev. D \textbf{63}
(2001) 041301.

\bibitem{5} G. Huey, J. E. Lidsey, Phys. Lett. B\textbf{514} (2001) 217.

\bibitem{6} L. H. Ford, Phys. Rev. D\textbf{35} (1987) 2955.

\bibitem{7} B. Feng, M. Li, Phys. Lett. B\textbf{564} (2003) 169.

\bibitem{8} A. R. Liddle, L. A. Urena-Lopez, Phys. Rev. D\textbf{68} (2003)
043517.

\bibitem{9} M. Sami, V. Sahni, Phys. Rev. D\textbf{70} (2004) 083513.

\bibitem{deffayet} C. Deffayet, G. R. Dvali, G. Gabadadze, A. I. Vainshtein,
%"Nonperturbative continuity in graviton mass versus perturbative discontinuity",
Phys. Rev. D\textbf{65} (2002) 044026 [hep-th/0106001]; A. Nicolis, R.
Rattazzi, %"Classical and quantum consistency of the DGP model",
JHEP \textbf{0406} (2004) 059, [hep-th/0404159].

\bibitem{quiros} I. Quiros, R. Garc\'{\i}a-Salcedo, T. Matos, C. Moreno,
%"Self-interacting scalar field trapped in a DGP brane: The dynamical systems perspective"
Phys. Lett. B \textbf{670} (2009) 259--265.

\bibitem{speedlimit} E. Silverstein, D. Tong, Phys. Rev. D \textbf{70}
(2004) 103505; M. Alishahiha, E. Silverstein, D. Tong, Phys. Rev. D \textbf{%
70} (2004) 123505.

\bibitem{chen} X. Chen, %"'Multi-throat brane inflation"',
Phys. Rev. D \textbf{71} (2005) 063506 [hep-th/0408084]; JHEP \textbf{08}
(2005) 045 [hep-th/0501184].

\bibitem{basic} Z-K. Guo, N. Ohta,
%"Cosmological evolution of the Dirac-Born-Infeld field",
JCAP \textbf{04} (2008) 035.

\bibitem{DBInflation} M. C. Bento, O. Bertolami, A. A. Sen, Phys. Rev. D
\textbf{67} (2003) 063511; X. Chen, Phys. Rev. D \textbf{71} (2005) 063506;
S. E. Shandera, S-H. H. Tye, JCAP \textbf{05} (2006) 007.

\bibitem{chen1} X. Chen, M. Huang, S. Kachru, G. Shiu, JCAP \textbf{01}
(2007) 002.

\bibitem{dvali} G. R. Dvali, S. H. H. Tye, Phys. Lett. B \textbf{450} (1999)
72 [hep-ph/9812483]. S. Kachru, R. Kallosh, A. Linde, J. M. Maldacena, L.
McAllister, S. P. Trivedi, JCAP \textbf{0310} (2003) 013 [hep-th/0308055].

\bibitem{verlinde} H. Verlinde, Nucl. Phys. B \textbf{580} (2000) 264
[hep-th/9906182]; S. Gukov, C. Vafa, E. Witten,
%"CFT's from Calabi–Yau four-folds",
Nucl. Phys. B \textbf{584} (2000) 69 [hep-th/9906070]; S. Gukov S, C. Vafa,
E. Witten E, Nucl. Phys. B \textbf{608} (2001) 477 (erratum); K. Dasgupta,
G. Rajesh, S. Sethi, %"M theory, orientifolds and G-flux",
JHEP \textbf{08}(1999) 023 [hep-th/9908088]; B. R. Greene, K. Schalm, G.
Shiu, %"Warped compactifications in M and F theory",
Nucl. Phys. B \textbf{584} (2000) 480 [hep-th/0004103]; S. B. Giddings, S.
Kachru, J. Polchinski,
%"Hierarchies from fluxes in string compactifications",
Phys. Rev. D \textbf{66} (2002) 106006 [hep-th/0105097].

\bibitem{string} J. M. Cline, %"String cosmology",
hep-th/0612129; S. H. Henry Tye, Lect. Notes Phys. \textbf{737} (2008)
949-974 [hep-th/0610221]. L. McAllister, E. Silverstein, Gen. Rel. Grav.
\textbf{40} (2008) 565-605 (2008) [arXiv:0710.2951].

\bibitem{maiden} S. Kecskemeti, J. Maiden, G. Shiu, B. Underwood,  JHEP \textbf{0609}, (2006), 076, [arXiv:hep-th/0605189]

\bibitem{coley} A. A. Coley, \textit{Dynamical systems and cosmology},
Dordrecht-Kluwer, Netherlands (2003).

\bibitem{deformedtaquion} I Quiros, T Gonzalez, D Gonzalez, Y Napoles, R Garcia-Salcedo and C Moreno, %"Dirac-Born-Infeld/Tachyon Fields Equivalence And Tachyon Dynamics",
    [arXiv:0709.2399]

\bibitem{chinos} W. Fang, Y. Li, K. Zhang, H.-Q. Lu, Class. Quant. Grav. \textbf{26} (2009), 155005, [arXiv:0810.4193].

\bibitem{koyama} K. Koyama, %"Ghosts in the self-accelerating universe",
Class. Quantum Grav. \textbf{24} (2007) R231 [arXiv:0709.2399].

\bibitem{Ahn}  C. Ahn, C. Kim, E. V. Linder, Phys. Rev. D\textbf{80}, (2009), 123016, [arXiv:0909.2637].

\bibitem{Copeland2} E. J. Copeland, S. Mizuno, M. Shaeri, [arXiv:1003.2881]

\bibitem{Koyama} K. Koyama, Class. Quantum Grav. \textbf{24} (2007) R231, [arXiv:0709.2399v2]

\bibitem{plb09} T. Gonzalez, T. Matos, I. Quiros, A. Vazquez-Gonzalez,
%"Self-interacting scalar field trapped in a Randall�Sundrum braneworld: The dynamical systems perspective"
Phys. Lett. B \textbf{676} (2009) 161--167.

\bibitem{quiros} I. Quiros, T. Gonzalez, D. Gonzalez, Y. Napoles, R. Garcia-Salcedo, C. Moreno, %"Dirac-Born-Infeld/Tachyon Fields Equivalence And Tachyon Dynamics ",
[arXiv:0906.2617].

\end{thebibliography}
\end{document}